\documentclass[sigconf]{acmart}

\makeatletter                   
\def\mdseries@tt{m}             
\makeatother
\usepackage[draft=true]{minted} 
\usepackage{color}
\usepackage{hyperref}           
\usepackage{breakurl}           
\usepackage{booktabs} 
\usepackage{url}
\usepackage{graphicx}
\usepackage{listings}
\usepackage{algorithm}
\usepackage{algpseudocode}
\usepackage{subcaption}
\usepackage{enumitem}
\usepackage{syntax}
\usepackage{amsmath}
\usepackage{bigstrut}
\usepackage{mathpartir}
\usepackage{cleveref}
\usepackage{adjustbox}
\usepackage{nccmath}
\usepackage{bbding}

\crefformat{section}{\S#2#1#3}
\crefformat{subsection}{\S#2#1#3}
\crefformat{subsubsection}{\S#2#1#3}

\algrenewcommand\algorithmicrequire{\textbf{Input:}}
\algrenewcommand\algorithmicensure{\textbf{Output:}}

\setlist[itemize]{noitemsep, leftmargin=1em}
\setlist[enumerate]{noitemsep, leftmargin=1em}

\setcopyright{rightsretained}
\setlist[itemize]{noitemsep, leftmargin=1em}
\setlist[enumerate]{noitemsep, leftmargin=1em}

\newcommand{\prim}[1]{\text{{\sffamily{\small {\bf #1}}}}}
\newcommand{\primsem}[1]{\text{{\sffamily{\footnotesize {\bf #1}}}}}

\definecolor{dkred}{RGB}{87,10,10}

\lstdefinelanguage{puppet}{
  morecomment = [l]{;},
  morestring=[b]",
  sensitive = true,
  classoffset=0,
  morecomment=[l]\#,
  morekeywords={
      service,
      file,
      package,
      if, 
      class,
      define,
      exec,
  },
  classoffset=1, keywordstyle=\ttfamily\color[rgb]{0,0,1},
  morekeywords={
      path,
      ensure,
      name,
      content,
      require,
      subscribe,
      File,
      mode,
      group,
      owner,
      provider,
      source,
      command,
      creates,
      export,
      unless,
      replace,
      true,
      false,
  },
  alsoletter={\%},
  keywordsprefix={\%},
}

\lstdefinelanguage{strace}{
  morestring=[b]",
  sensitive = true,
  classoffset=0,
  classoffset=1,
  keywordstyle=\ttfamily\color[rgb]{0.627,0.126,0.941},
  stringstyle=\ttfamily\color[rgb]{0,0,1},
  morekeywords={
      |,
      O_RDONLY,
      O_RDWR,
      O_CREAT,
      NULL,
      PROT_READ,
      PROT_EXEC,
      MAP_PRIVATE,
      MAP_DENYWRITE,
      CLONE_CHILD_CLEARTID,
      CLONE_CHILD_SETTID,
      SIGCHLD,
  },
  alsoletter={\%},
  keywordsprefix={\%},
}

\setcopyright{none}
\renewcommand\footnotetextcopyrightpermission[1]{}

\begin{document}
\sloppy
\title{Detecting Missing Dependencies and Notifiers in Puppet Programs}

\author{Thodoris Sotiropoulos}
\email{theosotr@aueb.gr}
\affiliation{Athens University of Economics and Business}

\author{Dimitris Mitropoulos}
\email{dimitro@aueb.gr}
\affiliation{Athens University of Economics and Business}

\author{Diomidis Spinellis}
\email{dds@aueb.gr}
\affiliation{Athens University of Economics and Business}

\begin{abstract}

Puppet is a popular computer system
configuration management tool.
It provides abstractions
that enable administrators to setup
their computer systems declaratively.
Its use suffers from two potential pitfalls.
First,
if ordering constraints are not specified
whenever an abstraction depends on another,
the non-deterministic application of abstractions
can lead to race conditions.
Second,
if a service is not tied to its resources
through notification constructs,
the system may operate in a stale state
whenever a resource gets modified.
Such faults can degrade a computing
infrastructure's availability
and functionality.

We have developed an approach
that identifies these issues
through the analysis of a Puppet program
and its system call trace.
Specifically,
we present a formal model for traces,
which allows us to capture
the interactions of Puppet abstractions
with the file system.
By analyzing these interactions
we identify (1) abstractions
that are related to each other
(e.g., operate on the same file),
and (2) abstractions that should act as notifiers
so that changes are correctly propagated.
We then check the relationships from the trace's analysis
against the program's dependency graph:
a representation containing all the ordering constraints
and notifications declared in the program.
If a mismatch is detected,
our system reports a potential fault.

We have evaluated our method
on a large set of Puppet modules,
and discovered 57 previously unknown issues in 30 of them.
Benchmarking further shows
that our approach can analyze in minutes real-world configurations
with a magnitude measured in thousands of lines
and millions of system calls.
\end{abstract}



\maketitle

\section{Introduction}
\label{sec:intro}

The prevalence of cloud computing
and the advent of microservices
have made the management of
multiple deployment and testing
environments a challenging and
time-consuming task~\cite{iac, DJV10, PW16, dds}.
\emph{Infrastructure as Code} (IaC)
methods and tools automate
the setup and provision of these environments,
promoting reliability, documentation, and reuse~\cite{dds}.
Specifically,
IaC (1) boosts the reliability
of an infrastructure,
because it minimizes the human
intervention which is both time-consuming
and error-prone;
(2) ensures the predictability
and consistency of the final product,
because it eases the repetition
of the steps followed
to produce a specific outcome; and
(3) allows the documentation and reuse of a system's
configuration,
because it associates the system's configuration with
modular code~\cite{iac,automation,building,dds}.

Puppet~\cite{Loo11}
is one of the most popular system
configuration tools used in the IaC
context~\cite{rehearsal,security}.
Puppet abstracts the actual system resources
through a declarative approach.
It collects all the declared abstractions from a program,
and applies them one-by-one
so that the system eventually reaches the desired state.

By default,
any execution sequence of abstractions is valid,
unless there are specific ordering constraints
imposed by the inter-dependencies among them.
For example,
an Apache service should run only
after the installation of the corresponding package.
Therefore,
developers need to declare
any ordering constraints between abstractions
in their programs to remove erroneous execution sequences,
e.g., trying to start a service
before the installation of its package.
Conceptually,
Puppet captures all the ordering relationships
defined in a program through a directed acyclic graph
and applies each abstraction in topological ordering.
In this context,
all the unrelated abstractions are processed
\emph{non-deterministically}.
Furthermore,
Puppet allows programmers
to apply certain abstractions
whenever specific events take place
via a feature called \emph{notification}.
Notifications propagate changes to related resources,
ensuring that their state is up-to-date.
For instance,
when a configuration file changes
the corresponding service has to be notified
so that it will run with the new settings.

Tracking all the required ordering constraints
and notifications is a complicated task though,
mostly because
developers are not always aware of
the actual interactions of Puppet abstractions
with the underlying operating system.
Notably,
such errors can have a negative impact on
the reliability of an organization's infrastructure
leading to inconsistencies~\cite{rehearsal}
and outages~\cite{github-usecase}.
For example,
the Github's services
became unavailable
when a missing notifier in
their Puppet codebase
caused a chain of failures
such as {\sc dns} timeouts~\cite{github-usecase}.

Approaches that automatically
detect these issues in production code~\cite{rehearsal,citac}
have significant room for improvement,
facing limitations that prevent them from being practical.
{\it Rehearsal}~\cite{rehearsal}
employs static code verification
and cannot handle realistic Puppet programs.
In particular,
it cannot reason about programs
that abstract arbitrary shell commands.
Additionally,
the model-based testing approach adopted by
{\it Citac}~\cite{citac} imposes a
significant overhead 
and restrictions on
the supported Puppet programs under test
(they must be able to run in
Docker\footnote{\url{https://www.docker.com/}}
containers).
It also requires extra instrumentation
on the execution engine of Puppet.
Finally,
none of those tools
addresses missing notification issues.

We have developed a {\it practical} and
{\it effective} approach to identify
faults involving ordering violations
and notifiers in Puppet programs.
To do so,
we examine the system call trace produced
by a single execution.
The stepping stone of our approach is
\emph{FStrace};
a language for modeling
a sequence of system call traces. 
We employ FStrace
and operate in the following steps.
First,
we {\it model} the system call trace
of a Puppet program in FStrace.
Through FStrace,
we derive an analysis
that {\it captures} the interactions of
higher-level programming constructs
(Puppet abstractions)
with the file system,
and {\it estimates} the set of the expected
relationships among them.
Then,
for a given Puppet program,
we {\it build} the \emph{dependency graph}
which reflects all the ordering relationships
and notifications that have been specified by the developer.
Finally,
we {\it verify} whether the expected relationships
(as specified from the analysis of traces)
hold with respect to the dependency graph.
Unlike previous tools~\cite{rehearsal,citac},
our approach (1) can reason about
which system resources are affected by
the execution of a program and how,
and (2) requires only a single Puppet run
for discovering issues.\\
\vspace{0.5mm}
{\bf Contributions.}
Our work makes the following contributions:
\vspace{-0.7mm}
\begin{itemize}
\item We introduce FStrace,
a language for modeling system call traces.
The interpretation rules of FStrace
allow us to infer the impact
that higher-level building blocks have on the file system.
The model proposed is generic
and can be leveraged---apart from Puppet programs---by
other domains (Section~\ref{sec:modeling}).

\item We design a framework for detecting
faults regarding ordering violations
and notifiers in Puppet programs.
To the best of our knowledge,
it is the first approach to deal with issues
involving notifiers.
(Section~\ref{sec:faults}).

\item We provide an open-source implementation of our approach
(Section~\ref{sec:implementation}).

\item We demonstrate the effectiveness
and performance of our tool
on a large set of Puppet modules.
Specifically,
our tool was able to detect 57 previously
unknown faults in 30 Puppet modules.
We provided fixes for 21 projects and
16 of them were accepted and integrated.
This implies that our tool is capable
of discovering issues
that are useful to developers
(Section~\ref{sec:evaluation}).

\end{itemize}

\section{Overview}
\label{sec:overview}

Here is a brief overview of Puppet,
a motivating example
that demonstrates the types of defects
our approach detects, and how our approach
is structured.

{\bf Puppet.}
Puppet enables developers
to describe the desired state of a
system declaratively.
\begin{lstlisting}
package {"apache2": ensure => "installed"}
file {"/etc/apache2/apache2.conf": ensure => "file"}
service {"apache2": ensure => "running"}
\end{lstlisting}
The code above indicates that the
{\tt apache2} package should be installed in the host,
the file {\tt apache2.conf}
should exist in the {\tt /etc/apache2/} path,
and that the Apache server should be running.
There are different types for abstracting
system resources,
including but not limited to,
{\tt file},
{\tt package},
{\tt service},
{\tt exec}.
Beyond that,
the Puppet language
provides conditionals,
loops, and---for reusability---supports
the creation of new abstractions and classes.

Puppet code is stored in files called \emph{manifests}.
Puppet compiles manifests into \emph{catalogs}
that specify all the abstractions
it needs to apply
in a particular system to reach
the desired state~\cite{catalog-compilation}.
Then,
it evaluates the compiled catalogs
and applies potential changes,
if the system is not in the appropriate state.
For example,
if a file does not exist at a certain location,
Puppet will create it.
The execution of a catalog must be
\emph{idempotent}~\cite{idempotence},
so that the evaluation proceeds only
if the current and the desired state of
the system do not match.

{\bf Motivating Example.}
In the following,
we present a motivating example
that demonstrates the issues
that our approach addresses.

\begin{figure}[t]
\begin{lstlisting}[language=puppet]
class ntp (..., String $defaults_file = "/etc/default/ntp") {
  package { "ntp": ensure => $package_ensure }
  file { "/etc/ntp.conf":
    ensure  => "present",
    require => Package[$package],
  }
  if $defaults_file {
    file { $defaults_file:
      ensure  => "present",
      content => "conf content..."
      require => Package["ntp"],
    }
  }
  $service_subscribe = $defaults_file ? {
    true    => [ File[$defaults_file], File["/etc/ntp.conf"] ],
    default => [ File["/etc/ntp.conf"] ],
  }
  service { "ntp":
    ensure     => "running",
    require    => File[/etc/ntp.conf],
    subscribe  => $service_subscribe,
  }}
\end{lstlisting}
\vspace{-4mm}
\caption{A Puppet program that
manifests a missing ordering relationship and notifier.
We omit irrelevant code.}
\label{fig:motivation-ex1}
\vspace{-4.5mm}
\end{figure}

\emph{Missing Ordering Relationships (MOR)}
occur when a developer fails to define
a {\it happens-before} relation
between two Puppet abstractions
that depend on each other.
This can lead to unstable code
that behaves correctly in some circumstances,
but breaks in others
depending on the order that
Puppet processes resources.

Consider the code snippet shown in Figure~\ref{fig:motivation-ex1}.
This fragment is taken from a real-world
Puppet module~\cite{ntp},
which defines a class
that setups a Network Time Protocol ({\sc ntp}) service.
This class expects 
the {\tt String} parameter {\tt \$defaults_file}
as an argument (line 1),
which stands for the path
where the service's default configuration file is created.
Notice that
the default value of {\tt \$defaults_file}
is {\tt /etc/default/ntp}.
Initially,
at line 2,
the program installs the {\tt ntp} package.
Then,
it creates a configuration file
at the path {\tt /etc/ntp.conf} (lines 3--6).
If the variable {\tt \$defaults_file} is \emph{defined},
Puppet generates a file
at the location specified by {\tt \$defaults_file}
(lines 7--13).
Note that if Puppet finds an already defined
variable in the condition of an {\tt if} statement,
it implicitly coerces it to {\tt true}.
Puppet evaluates both {\tt file} abstractions
\emph{after} the {\tt package} resource.
This is expressed through
the {\tt require} property at lines 5, 11.

In lines 18 to 22,
the program declares that an {\sc ntp} service
should be in a running state.
Notice the {\tt subscribe} parameter at line 21,
where the {\sc ntp} service subscribes to
the variable {\tt \$service_subscribe}
which in turn is computed at lines 14--17.
The {\tt subscribe} construct
states that the {\tt service} depends on
the Puppet abstractions included
in the {\tt service_subscribe} variable.

The initial intention of the programmer is
that when the {\tt \$defaults_file} is defined,
then the service should subscribe to
both {\tt File[\$default_file]}
and {\tt File["/etc/ntp.conf"]} (line 15).
However,
unlike {\tt if} statements,
the operand of the ``?'' operator
(i.e, {\tt \$defaults_file})
at line 14 \emph{never} evaluates to {\tt true}
because it is a {\tt String} variable.
Therefore,
the program considers only the default case (line 16)
where the subscribers' list only contains
{\tt File[/etc/ntp.conf]}.
As a result,
the dependency between the {\tt service}
and the {\tt /etc/default/ntp} file is never created.
In other words,
Puppet might apply {\tt service}
before the configuration of {\tt /etc/default/ntp}.
A fix to this problem is to replace ``?'' operator
with an {\tt if} statement.

\emph{Missing Notifiers (MN).}
Notifiers are necessary for many
entities such as configuration files and services.
An update to a configuration file should trigger
the restart of the corresponding service,
because these files typically
describe settings processed during
a service's startup.
For example the configuration file
of an Apache service lists additional modules
that should be loaded into memory.

A missing notifier issue is illustrated
in Figure~\ref{fig:motivation-ex1}.
The {\tt subscribe} primitive (line 21)
creates a notifier
that restarts the {\sc ntp} service
whenever there is a change to the resources
included in the {\tt \$service_subscribe} list.
Although the programmer's intention was correct,
the programming error at lines 14--17
causes an unexpected behavior:
the service does not restart
even if the configuration {\tt /etc/default/ntp} changes
because the service subscribes only to {\tt /etc/ntp.conf}.

\begin{figure}[t]
  \begin{center}
    \includegraphics[scale=0.45]{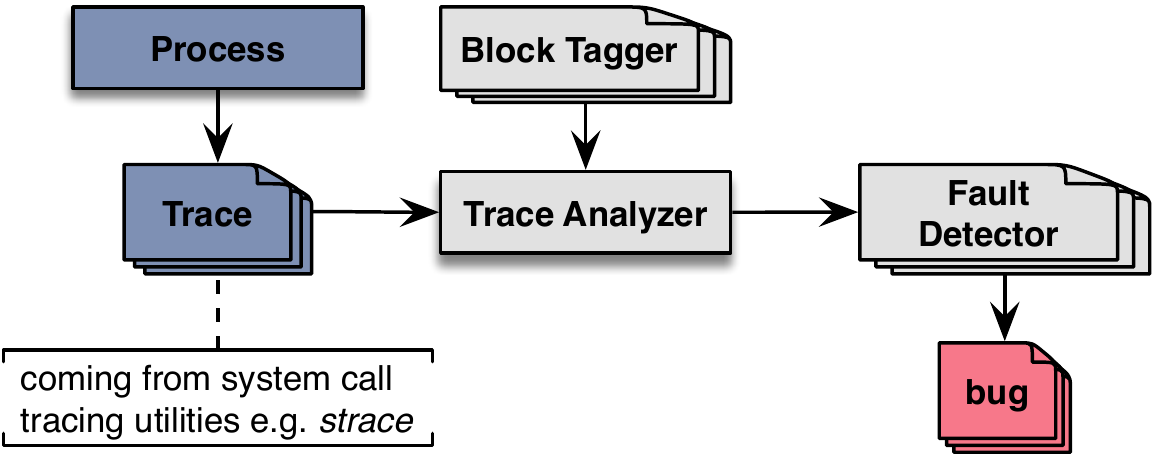}
  \end{center}
  \vspace{-3mm}
  \caption{The Abstract Architecture of our Framework.}
  \label{fig:arch}
  \vspace{-6mm}
\end{figure}

\begin{figure*}[t]
\begin{lstlisting}[language=strace]
...
103   write(1, "Info: /Stage[main]/Ntp/File[/etc/default/ntp]: Starting to evaluate the resource", 91) = 91
103   open("/etc/default/ntp20190128-32-15kba2r", O_RDWR|O_CREAT, 0600) = 7
103   write(7, "conf content"..., 44)   = 44
103   close(7)                          = 0
103   rename("/etc/default/ntp20190128-32-15kba2r", "/etc/default/ntp") = 0
103   write(1, "Info: /Stage[main]/Ntp/File[/etc/default/ntp]: Evaluated in 0.06 seconds", 83) = 83
103   write(1, "Info: /Stage[main]/Ntp/Service[ntp]: Starting to evaluate the resource", 81) = 81
...
103   execve("/etc/init.d/ntp", ["/etc/init.d/ntp", "start"], ...) = 0
...
650   clone(child_stack=NULL, flags=CLONE_CHILD_CLEARTID|CLONE_CHILD_SETTID|SIGCHLD, child_tidptr=0x7f70159c39d0) = 660
...
660   open("/etc/default/ntp", O_RDONLY) = 3
660   read(3, "conf content..."..., 44) = 44
103   write(1, "Info: /Stage[main]/Ntp/Service[ntp]: Evaluated in 1.85 seconds", 73) = 73
\end{lstlisting}
  \vspace{-2mm}
\caption{An example of trace produced
by the {\tt strace}.}\label{fig:strace-ex}
\end{figure*}

{\bf Framework.}
To address these issues,
we propose a framework---illustrated in
Figure~\ref{fig:arch}---that operates as follows.
First,
it monitors the system calls of the Puppet process and its descendants.
Then, the framework employs two components:
the \emph{trace analyzer},
and the \emph{fault detector}.
The trace analyzer takes as input a system
cal trace derived from the application
of a Puppet configuration,
and it interprets each system call
based on the model
described in Section~\ref{sec:modeling}.
The analyzer is instantiated
with the \emph{block tagger} component,
which splits system calls into different blocks
that correspond to Puppet abstractions.
The analysis output is the set of the effects
that Puppet abstractions have on the file system.
The fault detector,
generates the directed acyclic graph containing
all the ordering constraints and
notifications declared in a Puppet program,
and compares it against the expected relationships
inferred from the output of the trace analysis.
If a mismatch is identified,
the fault detector reports a potential fault.

{\bf Trace Example.}
To generates traces,
we exploit a system call tracing
program~\cite{Rod86,MMG06},
namely,
{\tt strace}.
Figure~\ref{fig:strace-ex} presents
an excerpt from the trace of the program
of Figure~\ref{fig:motivation-ex1}.
Each line denotes
an invocation of a system call
along with the process ({\sc pid})
that triggered it.
For example,
the entry {\tt 103 close(7) = 0} states
that the process with {\sc id} = 103,
invoked {\tt close} with 7 as an argument,
and that system call returned 0.
By further inspecting
Figure~\ref{fig:strace-ex},
we observe that
Puppet initially processes
the {\tt File[/etc/default/ntp]} resource (lines 2--7),
and then the {\sc ntp} service (lines 8--16).
Observe the calls of {\tt write} at lines 2, 7--8, 16.
These calls correspond to messages
printed to the standard output by Puppet
indicating the points
where the application of each abstraction
starts and ends respectively.
We exploit these points to classify
system calls according to the Puppet abstraction
they come from (Section~\ref{sec:puppet-modeling}).

\vspace{-2mm}

\section{Modeling System Call Traces}
\label{sec:modeling}

We formally introduce \emph{FStrace},
the language we use
to model system calls in it,
and
discuss how we model traces
stemming from Puppet programs.

\subsection{The FStrace Language}
\label{sec:language}

\begin{figure}[h]
  \centering
  \begin{fleqn}
  \[
      f \in F = \mathbb Z,\:\:
      z \in Proc = \mathbb Z^{*},\:\:
      b \in BlockID,\:\:
      v \in Filename
  \]
  \end{fleqn}
  \vspace{-5mm}
  \begin{grammar}
      <$e \in Trace$> ::= $x^{*}$

      <$x \in Block$> ::= \prim{begin} $b$ $(z, s)^{*}$ \prim{end} 

      <$s \in Sys$> ::= \prim{chdir} $p$ |\
      \prim{clone} $c^{*}$ $f$ |\
      \prim{close} $f$ |\
      \prim{dupfd} $f_1$ $f_2$ |\\
      \prim{hpath} $d$ $p$ $m$ |\
      \prim{hpathsym} $d$ $p$ $m$ |\
      \prim{link} $d_1$ $p_1$ $d_2$ $p_2$ |\\
      \prim{open} $d$ $p$ $o^{*}$ $f$ |\
      \prim{rename} $d_1$ $p_1$ $d_2$ $p_2$ |\
      \prim{symlink} $d$ $p_1$ $p_2$ |
      \prim{nop}

      <$m \in Eff$> ::= \prim{consumed} |\
      \prim{produced} |\
      \prim{expunged}

      <$c \in CloneFlags$> ::= \prim{fd} |\
      \prim{cwd}

      <$m \in OpenFlags$> ::= \prim{read} |\
      \prim{write} |\
      \prim{trunc} |\
      \prim{creat}

      <$d \in DirFd$> ::= $f$ |\ \prim{at\_fdcwd}

      <$p \in Path$> ::= $v^{*}$
  \end{grammar}
  \vspace{-4mm}
  \caption{The syntax of FStrace.}\label{fig:syntax}
  \vspace{-4mm}
\end{figure}

FStrace primitives are designed to
model system calls
that operate on file system resources.
Some of the constructs have
direct correspondence with the actual system calls,
while others are generic enough
so that they can represent a family of system calls.
We group system calls into execution blocks,
with a unique {\sc id}.
FStrace assumes
that within a block,
all system calls are processed sequentially.
However,
the execution order at the level of blocks
is \emph{not} deterministic.
Therefore,
there is no guaarantee
that a block $b_1$ is always processed before $b_2$,
even if the former appears before the latter in traces.
FStrace processes every execution block \emph{atomically},
and nested blocks are not allowed.

\subsubsection{Syntax and Domains}
\label{sec:snd}

Figure~\ref{fig:syntax} shows
the syntax of FStrace.
The language constists of
file names,
paths---which are sequences of file names---and
file descriptors represented by
either an integer or the \prim{at\_fdcwd} construct.
We also include (1) flags
(e.g., read-mode, write-mode, \prim{o\_trunc})
that indicate how a file is opened,
(2) the constructs \prim{fd} and \prim{cwd}
that provide information for cloning a process,
(3) primitives (\prim{consumed}, \prim{produced}, \prim{expunged}),
that stand for the types of the effect
that a system call has on a file,
and (4) an infinite set of
unique identifiers for execution blocks.
A trace is a sequence of blocks.
A block has the following syntax:
\prim{begin} $b$ $(z, s)^{*}$ \prim{end},
where $b$ implies its {\sc id}
and $(z, s)^{*}$ is a sequence of trace entries.
Each pair $(z, s)$ is a process {\sc id} ({\sc pid}),
which is a positive integer,
and a system call.
Finally,
FStrace models every system call $s \in Sys$
using the following eleven constructs.
\\
\vspace{-1.8mm}

\noindent
\prim{chdir} $p$
changes the working directory of
the current process to $p$.\\ 
\noindent
\prim{clone} $c^{*}$ $f$
spawns
a new process whose {\sc pid} is $f$.
The given flags $c^{*}$ reveal
what kind of information is shared
between the parent and the child process.
\\
\noindent
\prim{close} $f$
disassociates the file descriptor $f$ from the corresponding resource.\\
\noindent
\prim{dupfd} $f_1$ $f_2$
creates a new file descriptor $f_2$
as a copy of $f_1$.
This construct models
a number of system calls such as
{\tt dup}, {\tt dup2}, {\tt dup3}, {\tt fcntl(fd, F\_DUPFD)}.\\
\noindent
\prim{hpath} $d$ $p$ $m$
captures the effect $m$
that a system call has on the path $p$.
If the given path name is not absolute,
we interpret it as relative to the file descriptor $d$.
If the value of $d$ is \prim{at\_fdcwd},
we consider $p$ relative to
the current working directory.
Otherwise,
if $p$ is absolute,
we ignore $d$.
In this way,
we can represent the system calls
whose suffix is \emph{``at''}
(e.g., {\tt linkat}, {\tt renameat})
or system calls that take relative paths as arguments.
For instance,
the system call {\tt stat("foo",}\dots{\tt)}---which
retrieves the main information
and attributes of the file {\tt foo}---is
represented as \prim{hpath at\_fdcwd} {\tt foo} \prim{consumed}.
On the other hand,
we represent the system call {\tt mkdir("/foo/bar")}---which
creates a new directory at path {\tt /foo/bar}---as
\prim{hpath at\_fdcwd} (``/'', ``foo'', ``bar'') \prim{produced}.\\
\noindent
\prim{hpathsym} $d$ $p$ $m$
operates
in a way similar to \prim{hpath}.
In \prim{hpathsym} though,
if the given path $p$ is a symbolic link,
we do not dereference it.
Through \prim{hpathsym}
we express system calls that do not follow
symbolic links such as
{\tt lstat}, {\tt lchown}, {\tt lgetxattr}.\\
\noindent
\prim{link} $d_1$ $p_1$ $d_2$ $p_2$
creates a hard link
that points to the same resource as
the file indicated by
the file descriptor $d_1$
and the path $p_1$. \\
\noindent
\prim{open} $d$ $p$ $o^{*}$ $f$
associates the file
indicated by the path $p$ with
the file descriptor $f$.
A sequence of flags $o^{*}$
captures the operations that can be performed on the file.\\
\noindent
\prim{rename} $d_1$ $p_1$ $d_2$ $p_2$
arranges that
a given file,
specified through the path defined by
the file descriptor $d_1$
and path $p_1$,
is accessed
through the new path defined by
$d_2$ and $p_2$. \\
\noindent
\prim{symlink} $p_1$ $d$ $p_2$
creates a symbolic link file at
the location specified by
the file descriptor $d$ and path $p_2$
pointing to the path $p_1$.\\
\noindent
\prim{nop} (no operation)
does not affect the state.
We use \prim{nop} to model all system calls
that we do not need to take into account,
e.g., {\tt getpid}, {\tt sync}.\\

Figure~\ref{fig:domains} illustrates
the semantic domains of FStrace.
FStrace introduces six major components:
An {\tt inode} table $\tau \in INodeT$
is a map of a pair,
consisting of an {\tt inode} and a file name
to another {\tt inode}.
An {\tt inode} is a positive integer
that acts as the \emph{identifier} for
a certain file system resource.
Note that we also keep
the special {\tt inode} $\iota_r$
which corresponds to the {\tt inode}
of the root directory ``/''.
The {\tt inode} table mimics
the {\tt inode} structure implemented
in Unix-like operating systems.
In this context,
the first element of the key
is the {\tt inode} of the directory
where the file name exists.
For example,
the {\tt inode} of the {\tt /foo} file,
whose value is 3,
is stored as follows:
$[(\iota_r, \text{``foo''}) \rightarrow 3)]$.
A file descriptor table $\pi \in FdT$
maps an address
and a file descriptor to an {\tt inode}.
We use this component
to map open file descriptors of a process
to the resource they handle.
The $CwdT$ element maps
an address to an {\tt inode}.
That {\tt inode}
stands for the current working directory of a process.

Observe that
we do not use the {\sc pid}
found in the trace entries
as the key of the two definitions above.
Instead,
we have an indirection:
each process points to a pair of addresses
(e.g., see the domain $ProcT$).
The first element of the pair is the address
that stores the file descriptor table of the process.
The second element of the pair reflects
the address
where the current working directory of the process is located.
Therefore,
two different processes might share
the same file descriptor table or working directory.
For example,
in the following entries:
$[(z_1 \rightarrow (\alpha_1, \alpha_2)),\ z_2 \rightarrow (\alpha_1, \alpha_3)]$,
the processes $z_1$ and $z_2$ point to the same
file descriptor table
because the first elements of their pairs are identical
(i.e., $\alpha_1$).
Similarly,
since their second addresses do not match
(i.e, $\alpha_2 \ne \alpha_3$),
we presume that they do not share
the same working directory;
thus,
a change imposed by any process
does not affect the other one.

A table of symbolic links $\kappa \in SymT$
is a partial map of {\tt inode}s to paths.
This domain holds the path names
that symbolic links---identified by their {\tt inode}s---point to.
The last component of FStrace ($\rho \in Res$) maps
path names to an element of the power set of
blocks and effects.
Specifically,
this component tracks
where and how each path is accessed.
For example,
the entry $\texttt{/foo} \rightarrow \{(\prim{produced}, b_1), (\prim{consumed}, b_2)\}$
indicates that the path {\tt /foo}
is produced in the block $b_1$
and consumed in $b_2$.
The state $\langle\tau, \pi, \phi, \nu, \kappa, \rho\rangle$
is a tuple consisting of
the six components described above.

\begin{figure}[t]
\centering
\small
\begin{align*}
    \iota \in INode  &= \{\iota_i \mid i \in \mathbb Z^{*}\} \cup \{\iota_r\}\\
    \alpha \in Addr  &= \{\alpha_i \mid i \in \mathbb Z^{*}\} \\
    \tau \in INodeT  &= (INode \times Filename) \rightarrow INode \\
    \pi \in FdT      &= Addr \hookrightarrow (F \hookrightarrow INode) \\
    \nu \in ProcT    &= Proc \rightarrow (Addr \times Addr) \\
    \phi \in CwdT    &= Addr \hookrightarrow INode \\
    \kappa \in SymT  &= INode \hookrightarrow Path \\
    \rho \in Res     &= Path \hookrightarrow \mathcal{P}(Eff \times Block)
\end{align*}
\vspace{-6mm}
\caption{Semantic domains for FStrace.}\label{fig:domains}
\vspace{-4mm}
\end{figure}

\subsubsection{Preliminary Definitions}
\label{sec:definitions}

A number of specific operations apply to FStrace's domains.
The binary operator $::$
denotes the addition of an element to a set,
while $\downarrow_i$ manifests the
projection of the $i^{th}$ element.
Also,
we define the following auxilliary functions:

\begin{itemize}
\item $I(p, \tau)$: returns the {\tt inode}
to which the path $p$ points based on
the inode table $\tau$.

\item $P(\iota, \tau)$: returns the paths
that point to the {\tt inode} $\iota$
according to the inode table $\tau$.

\item $join(p_1, p_2)$: joins the two paths $p_1$ and $p_2$.

\item $dir(p)$ returns the parent directory of the path $p$.

\item $base(p)$ returns the base name of the path $p$.
\end{itemize}

\noindent
We also define the function $\mathrm{Ab}(d, p, l, r, \tau)$
which gives the absolute path
for a given path name $p$
and a file descriptor $d$
with regards to the provided
open file descriptors $l$
and the current working directory $r$ of a process.

\[
    \mathrm{Ab}(d, p, l, r, \tau) =
    \begin{cases}
        p , & \text{if } p\downarrow_1 = \text{``/''} \\
        join(P(r, \tau)\downarrow_1,\ p)  & \text{if } d = \prim{at\_fdcwd} \\
        join(P(l(d), \tau)\downarrow_1,\ p), & \text{otherwise}
    \end{cases}
\]

\noindent
Finally,
the function $Op(m)$ gives the effect
that the {\tt open} system call
has on a file
based on the flags $m$.
$Op$ is defined as:
\[
    \mathrm{Op}(m) =
    \begin{cases}
        \prim{produced}, & \text{if } (\prim{trunc} \in m \land \prim{write} \in m)
        \lor \prim{creat} \in m \\
        \prim{consumed}, & \text{otherwise}
    \end{cases}
\]

\subsubsection{Semantics}
\label{sec:semantics}

Figure~\ref{fig:semantics} shows
the semantics of FStrace.
We present a subset of our rules for brevity.
Each rule follows
the form below:

\[
    \langle\tau, \pi, \phi, \nu, \kappa, \rho\rangle \xrightarrow{b,e}
    \langle\tau', \pi', \phi', \nu', \kappa', \rho'\rangle
\]

\noindent
The relation $\xrightarrow{b,e}$ indicates
that given a trace entry $e$
(a pair of a {\sc pid} and a system call)
in execution block $b$,
the initial state $\langle\tau, \pi, \phi, \nu, \kappa, \rho\rangle$
transitions to
a new state $\langle\tau', \pi', \phi', \nu', \kappa', \rho'\rangle$.

\begin{figure*}
\centering
\footnotesize
\begin{mathpar}
\inferrule[chdir]{
      e = z, \primsem{chdir}\ p \\\\
      \alpha = \nu(z)\downarrow_2 \\
      \phi' = \phi[\alpha \rightarrow I(p, \tau)]
  }
  {\langle\tau, \pi, \phi, \nu, \kappa, \rho\rangle \xrightarrow{b,e} \langle\tau, \pi, \phi', \nu, \kappa, \rho\rangle}
\hva \and
\inferrule[clone-copy]{
      e = z, \primsem{clone}\ \emptyset\ f \\\\
      \text{fresh } \alpha_1 \\
      \text{fresh } \alpha_2 \\
      \nu' = \nu[f \rightarrow (\alpha_1, \alpha_2)] \\\\
      \pi' = \pi'[\alpha_1 \rightarrow \pi(z)] \\
      \phi'= \phi[\alpha_2 \rightarrow \phi(z)]
  }
  {\langle\tau, \pi, \phi, \nu, \kappa, \rho\rangle \xrightarrow{b,e} \langle\tau, \pi', \phi', \nu', \kappa, \rho\rangle}
\hva \and
  \inferrule[clone-share]{
      e = z, \primsem{clone}\ (\primsem{fd}, \primsem{cwd})\ f \\\\
      (\alpha_1, \alpha_2) = \nu(z) \\
      \nu' = \nu[f \rightarrow (\alpha_1, \alpha_2)]
  }
  {\langle\tau, \pi, \phi, \nu, \kappa, \rho\rangle \xrightarrow{b,e} \langle\tau, \pi, \phi, \nu', \kappa, \rho\rangle}
\hva \and
\inferrule[dupfd]{
      e = z, \primsem{dupfd}\ f_1\ f_2 \\\\
      \alpha = \nu(z)\downarrow_1 \\
      \iota = \pi(\alpha)(f_1) \\\\
      \pi' = \pi[(\alpha \rightarrow (\pi(\alpha)[f_2 \rightarrow \iota])]
  }
  {\langle\tau, \pi, \phi, \nu, \kappa, \rho\rangle \xrightarrow{b,e} \langle\tau, \pi', \phi, \nu, \kappa, \rho\rangle}
\hva \and
\inferrule[open]{
      e = z, \primsem{open}\ d\ p\ o\ f \\\\
      (\alpha_1, \alpha_2) = \nu(z) \\
      p' = \mathrm{Ab}(d, p, \pi(\alpha_1), \phi(\alpha_2), \tau) \\\\
      m = Op(o) \\
      \pi' = \pi[\alpha_1 \rightarrow \pi(\alpha_1)[f \rightarrow I(p')]] \\\\
      \rho' = \rho[p' \rightarrow (m, b) :: \rho(p')] 
  }
  {\langle\tau, \pi, \phi, \nu, \kappa, \rho\rangle \xrightarrow{b,e} \langle\tau, \pi', \phi, \nu, \kappa, \rho'\rangle}
\hva \and
\inferrule[hpath]{
      e = z, \primsem{hpath}\ d\ p\ m \\\\
      m \ne \primsem{expunged} \\
      p' = \mathrm{Ab}(d, p, \dots) \\\\
      p'' = \kappa(I(p')) \\
      p'' \neq {\bf undef} \\\\
      \rho' = \rho[p'' \rightarrow (m, b) :: \rho(p'')]
  }
  {\langle\tau, \pi, \phi, \nu, \kappa, \rho\rangle \xrightarrow{b,e} \langle\tau, \pi, \phi, \nu, \kappa, \rho'\rangle}
\hva \and
\inferrule[hpath-expng]{
      e = z, \primsem{hpath}\ d\ p\ \primsem{expunged} \\\\
      p' = \mathrm{Ab}(d, p, \dots) \\\\
      l = \{m\ \mid \forall m \in \rho(p'):\: m\downarrow_2 \neq b\}\\\\
      \rho' = \rho'[p' \rightarrow (\primsem{expunged}, b) :: l] \\\\
      p_1 = dir(p) \\
      p_2 = base(p) \\\\
      \tau' = \tau[(I(p_1), p_2) \rightarrow {\bf undef}]
  }
  {\langle\tau, \pi, \phi, \nu, \kappa, \rho\rangle \xrightarrow{b,e} \langle\tau', \pi, \phi, \nu, \kappa, \rho'\rangle}
\hva \and
\inferrule[link]{
      e = z, \primsem{link}\ d_1\ p_1\ d_2\ p_2 \\\\
      p_1' = \mathrm{Ab}(d_1, p_1, \dots) \\
      p_2' = \mathrm{Ab}(d_2, p_2, \dots) \\\\
      w = dir(p_2') \\
      t = base(p_2') \\\\
      \tau' = \tau[(I(w), t) \rightarrow I(p_1')] \\\\
      \rho' = \rho[p_2' \rightarrow (\primsem{produced}, b) :: \rho(p_2')]
  }
  {\langle\tau, \pi, \phi, \nu, \kappa, \rho\rangle \xrightarrow{b,e} \langle\tau', \pi, \phi, \nu, \kappa, \rho'\rangle}
\hva \and
\inferrule[symlink]{
      e = z, \primsem{symlink}\ p_1\ d\ p_2 \\\\
      p_2' = \mathrm{Ab}(d_2, p_2, \dots) \\
      \text{fresh } \iota \\\\
      w = dir(p_2') \\
      t = base(p_2') \\\\
      \tau' = \tau[(I(w), t) \rightarrow \iota] \\
      \kappa' = \kappa[\iota \rightarrow p_1] \\\\
      \rho' = \rho[p_2' \rightarrow (\primsem{produced}, b) :: \rho(p_2')]
  }
  {\langle\tau, \pi, \phi, \nu, \kappa, \rho\rangle \xrightarrow{b,e} \langle\tau', \pi, \phi, \nu, \kappa', \rho'\rangle}
\hva \and
\inferrule[rename]{
      e = z, \primsem{rename}\ d_1\ p_1\ d_2\ p_2 \\\\
      p_1' = \mathrm{Ab}(d_1, p_1, \dots) \\
      p_2' = \mathrm{Ab}(d_2, p_2, \dots) \\
      w_1 = dir(p_1') \\\\
      t_1 = base(p_1') \\
      w_2 = dir(p_2') \\
      t_2 = base(p_2') \\
      l = \rho(p_1') \\
      l' = \{e\ \mid \forall e \in l:\: e\downarrow_2 \neq \alpha\} \\\\
      \tau' = \tau[(I(w_2), t_2) \rightarrow I(p_1')][(I(w_1), t_1) \rightarrow {\bf undef}] \\
      \rho' = \rho[p_1' \rightarrow (\primsem{expunged}, b) :: l'][p_2' \rightarrow (\primsem{produced}, b) :: \rho(p_2')]
  }
  {\langle\tau, \pi, \phi, \nu, \kappa, \rho\rangle \xrightarrow{b,e} \langle\tau', \pi, \phi, \nu, \kappa, \rho'\rangle}
\end{mathpar}
\vspace{-3mm}
\caption{The interpretation rules of FStrace.}\label{fig:semantics}
\end{figure*}

{\tt [CHDIR]} changes
the working directory of the current process $z$.
First,
it inspects the process table
$\nu$ to get the address
that holds the value of the current
process's working directory.
Then,
it updates $\phi$
so that the address $\alpha$ points to
the {\tt inode} of the path $p$.

{\tt [CLONE-COPY]} demonstrates the case
where we spawn a new process $f$
by passing the empty sequence $c = \emptyset$.
In this case,
$f$ shares neither
the file descriptor table
nor the working directory
with the parent process $z$.
So,
we make copies of those values
by creating two fresh addresses
$\alpha_1, \alpha_2$.
Then,
we update the process table $\nu$ so that
the new process $f$ points to those new addresses.

{\tt [CLONE-SHARE]} behaves
in a similar way with {\tt [CLONE-COPY]}.
However,
this time,
the new process $f$ shares
the open file descriptors (flag \prim{fd})
and the working directory with $z$
(flag \prim{cwd}).
Therefore,
the freshly-created process $f$
points to the same addresses as $z$.

{\tt [DUPFD]} involves the scenario
where we duplicate a provided file descriptor.
Specifically,
we lookup the address $\alpha$ of
the current process's file descriptors table.
Then,
we retrieve the {\tt inode} $\iota$
pointed by the file descriptor $f_1$.
Finally,
we add the file descriptor $f_2$,
whose {\tt inode} value is $\iota$,
to the file descriptor table of $z$.

{\tt [OPEN]} opens a file
and returns a new file descriptor.
First,
it inspects the addresses $\alpha_1, \alpha_2$
where the file descriptor table
and the working directory of
the process $z$ are located.
Through the $\mathrm{Ab}$ function,
it computes the absolute path $p'$
using the file descriptor $d$,
and the path $p$.
This computation is boilerplate,
so we abbreviate it as $\mathrm{Ab}(d, p, \dots)$ in the next rules.
Given the flags $o^{*}$,
it estimates the effect $m$
that {\tt open} has on the path $p'$
(via the function $\mathrm{Op}$).
In turn,
it performs two updates.
First,
it adds $f$ to
the file descriptor table of the process $z$
using the address $\alpha_1$.
Notice that $f$ points to the {\tt inode} of the path $p'$
($f \rightarrow I(p')$).
Second,
it updates the $\rho$ element:
the path $p'$ receives the effect $m$
in the the block $b$.

{\tt [HPATH]} records the effect $m$
that a system call has in
the current execution block $b$.
It handles the case
when the given effect $m$ is not \prim{expunged}.
Specifically,
it determines the absolute path $p'$
through $\mathrm{Ab}(d, p, \dots)$.
It then inspects the symbolic link table to check
whether the path $p'$
points to another path or not.
If this is the case
(i.e., $\kappa(p') \ne {\bf undef}$),
we associate the resulting resource $p''$
with the effect $m$
in the current execution block $b$.
Note that
the \prim{hpathsym} operates similarly,
but it does not check
whether $p'$ is a symbolic link or not;
it just considers the path $p'$.

{\tt [HPATH-EXPNG]} illustrates the case
where we expunge the provided file.
As before,
we first compute the absolute path $p'$
associated with that resource.
Subsequently,
we remove all the effects
associated with $p'$
in the current execution block $b$,
leading to the set $l'$
(i.e., $l = \{m\ \mid \forall m \in \rho(p'):\: m\downarrow_2 \neq b\}$).
We add the \prim{expunged} effect to $l'$,
and finally,
we \emph{unlink} the path $p'$
from the {\tt inode} table.
For unlinking,
the pair $(I(p_1), p_2)$ refers to {\bf undef}.
Notice that $p_1$ is the parent directory of $p'$,
and $p_2$ is the base name of $p'$.

{\tt [LINK]} creates a hard link between two files.
As a starting point,
we take the absolute paths $p_1'$ and $p_2'$,
where $p_1'$ corresponds to the existing file,
while $p_2'$ stands for the path
where we create the hard link.
Then,
the {\tt inode} of the new path $p_2'$ is identical
to that of $p_1'$
($\tau' = \tau[(I(w), t) \rightarrow I(p_1')]$).
We also change the table $\rho$
so that the path $p_2'$ is produced
in the current execution block $b$.

{\tt [SYMLINK]} creates
a new symbolic link
that points to the path $p_1$.
It first estimates the absolute path $p_2'$
of the fresh symbolic link.
Then,
it creates a new {\tt inode} $\iota$
which the symbolic link points to
by updating the {\tt inode} table $\tau$.
It also changes the table $\kappa$
so that the new {\tt inode} $\iota$ targets
the path $p_1$.
Finally,
the path $p_2'$ is produced
in the current execution block $b$
leading to the new table $\rho'$.

{\tt [RENAME]} renames the name of a given file.
First,
it retrieves the absolute paths corresponding
to the old and the new path names
(i.e, $p_1'$ and $p_2'$).
Then,
it updates the {\tt inode} table $\tau$:
the {\tt inode} of $p_2'$ is the same with that of $p_1'$.
In turn,
it removes $p_1'$ from the inode table
(i.e., $p_1'$ points to {\bf undef}).
For these updates,
it is necessary to estimate:
(1) the {\tt inode} of their parent directories,
and (2) their base names.
Finally,
it updates the component $\rho$.
In particular,
it removes any effects on path $p_1'$
that took place within the block $b$,
and it marks $p_1'$ as \prim{expunged}
and $p_2'$ as \prim{produced} in $b$.

\subsection{Modeling Puppet Traces}
\label{sec:puppet-modeling}

To leverage FStrace we need to group system calls
into blocks
corresponding to higher-level programming constructs.
Specifically
for Puppet artifacts,
it makes sense to classify system calls
according to the Puppet abstraction
where they come from.
In this context,
we presume that
an execution block begins or ends
when the evaluation of a Puppet abstraction starts or terminates.
Thus,
the name of the execution block corresponds to
the name of the Puppet abstraction.

It is easy to identify the points
where the evaluation of a Puppet abstraction starts/finishes
by decoding the Puppet's debug messages.
Recall from Figure~\ref{fig:strace-ex}
that those messages appear
in the execution traces as writes to the standard output.
We have developed a block tagger for Puppet
that detects those debug messages
and marks them
as the entry and exit points of execution blocks.
For example,
consider again the traces in Figure~\ref{fig:strace-ex}.
We can model the trace entry at line 2
as the entry point of an execution block
whose name is ``{\tt File[/etc/default/ntp]}'',
whereas the system call at line 8
signals the ending of that execution block.
Hence,
all system calls
that appear between lines 2 and 8,
are included in the aforementioned block.

\vspace{-0.5mm}

\section{Detecting Faults}
\label{sec:faults}
\vspace{-0.5mm}

We locate faults in Puppet programs
by combining the trace analysis output
with the dependency graph:
a directed acyclic graph used to capture all
the ordering and notification relationships
between the abstractions of a given Puppet program.

\subsection{The Dependency Graph}

We consider the dependency graph as an element
of the following power set:
\vspace{-0.5mm}
$$
g \in DG = \mathcal{P}(P \times P \times L)
$$
\vspace{-0.5mm}
where $P$ is the set of Puppet abstractions,
and $L = \{\prim{notify}, \prim{before}\}$.
An entry $(p_1, p_2, l) \in g$ means
that the Puppet abstraction $p_2$ is dependent on $p_1$.
The label $l$ shows the relationship's type
between $p_1$ and $b_2$: if $l = \prim{before}$,
$p_1$ is processed before $p_2$,
whereas if $l = \prim{notify}$,
apart from preceding $p_2$,
$p_1$ also sends notifications to $p_2$.

\begin{figure}
\includegraphics[scale=0.5]{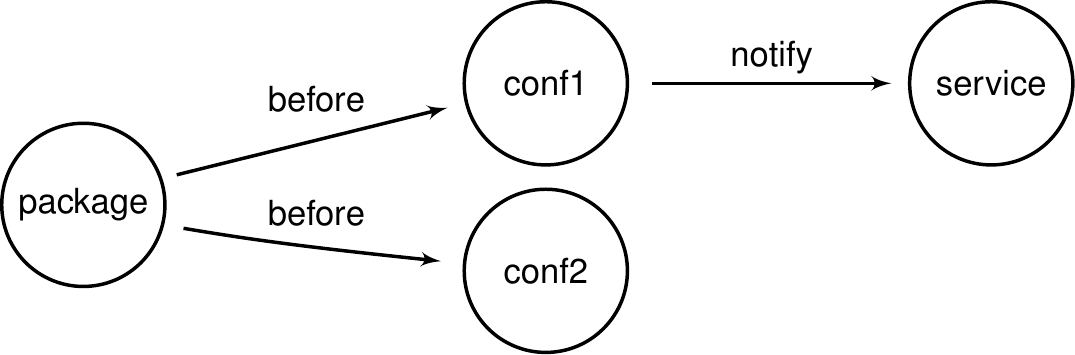}
\vspace{-2.5mm}
\caption{The dependency graph of the
program of Figure~\ref{fig:motivation-ex1}.}
\label{fig:graph}
\vspace{-4mm}
\end{figure}

Let the binary relation $\prec_g$
on nodes of a dependency graph $g \in CG$
defined as follows
$$p_1 \prec_g p_2 \Rightarrow (p_1, p_2, l) \in g,\
\text{where } p_1, p_2 \in P, l \in L$$
This relation is transitive so
$$
p_1 \prec_g p_2 \land p_2 \prec_g p_3 \Rightarrow p_1 \prec_g p_3,\
\text{where }, p_1, p_2, p_3 \in P
$$
The relation $\prec_g$ forms a \emph{happens-before} relation
between two Puppet abstractions
i.e., if there is path (of any length) from $p_1$ to $p_2$,
then the former is executed before the latter.

Similarly,
we define the transitive relation $\rightarrow_g$ on
nodes of a dependency graph $g \in CG$ as

\begin{align*}
p_1 \rightarrow_g p_2 &\Rightarrow (p_1, p_2, \prim{notify}) \in g\\
p_1 \rightarrow_g p_2 \land p_2 \rightarrow_g p_3 &\Rightarrow p_1 \rightarrow_g p_3,\
\text{where } p_1, p_2, p_3 \in P
\end{align*}
The relation $p_1 \rightarrow_g p_2$ indicates
that the abstraction $p_1$ notifies $p_2$.
For that purpose,
it only considers paths in $g$
with \prim{notify} edges.

Figure~\ref{fig:graph} depicts
the dependency graph of
the program of Figure~\ref{fig:motivation-ex1}.
For brevity,
the node {\tt conf1} stands for the {\tt /etc/ntp.conf},
and {\tt conf2} is the {\tt /etc/default/ntp} file.
We observe that the resource {\tt /etc/default/ntp}
has neither an ordering nor a notification relationship
with the {\tt service},
because the corresponding nodes are not connected to each other.
Also,
{\tt package} does not send notifications to {\tt service}
because there is no path from the former to the latter
where all edges have \prim{notify} labels.

\subsection{Combining FStrace
with the Dependency Graph}

\begin{algorithm}[t]
\small
\caption{Detecting Faults}
\label{alg:mor}
\begin{algorithmic}[1]
\Require $\rho \in Res, g \in CG$
\ForAll{$p$, $l$ {\bf in} $\rho$}
  \State get consumed $c = \{b \mid \forall (m, b) \in l.\ m = \prim{consumed}\}$
  \State get produced $t = \{b \mid \forall (m, b) \in l.\ m = \prim{produced}\}$
  \ForAll{$(b_1, b_2) \in t \times c$}
    \If{$b_1 \not\prec_{g} b_2 \land b_2 \not\prec_{g} b_1 \land b_1 \ne b_2$}
      \State report MOR between $b_1$, $b_2$ on path $p$
    \EndIf
    \If{\Call{isService}{$b_2$}}
      \If{$b_1 \not\rightarrow_{g} b_2$}
          \State report MN from $b_1$ to $b_2$
      \EndIf
    \EndIf
  \EndFor
\EndFor
\end{algorithmic}
\end{algorithm}

The dependency graph is a key element
in our fault detection approach.
Recall that the execution blocks in FStrace
are not totally ordered.
For example,
consider two blocks $b_1$, $b_2$
that affect the same file:
$b_1$ produces it and $b_2$
reads its contents.
In this case,
$b_2$ can be processed first,
because FStrace does not define
a temporal relation between the two blocks.
As a result,
there will be a failure
because $b_2$ will attempt to consume a file
that does not exist.

Thankfully,
the dependency graph of a Puppet program
can be employed to define the ordering
relationships of two execution blocks
(expressed through the $\prec_g$ relation).
Specifically,
we need to check
whether the $\prec_g$ relation is defined
for $b_1$, $b_2$ to identify missing ordering relationships.

For missing notifiers,
we first need to identify pairs of Puppet abstractions
where the application of the first element should
trigger the application of the second one.
To this end,
we look for blocks that
produce a particular resource $p$.
If the same resource $p$ is consumed by a block
that maps to a {\tt service},
presumably,
the blocks,
which produced $p$,
should have notification relationships
with the {\tt service} block.
That is,
if they produce an update to $p$,
{\tt service} should be refreshed to consume
the new version of $p$.

Algorithm~\ref{alg:mor} summarizes
our fault detection approach.
The algorithm expects as input
the map $\rho \in Res$---as specified
from the analysis of
traces---and a dependency graph $g \in CG$.
Then it iterates over every
key-value pair of $\rho$.
Recall that $\rho$ is a map
of a path $p$ to a set of pairs $l$;
each pair $(m, b) \in l$ stands for the effect $m$
that took place in the block $b$.
For a certain path $p$,
we retrieve the set of blocks $c$
that consumed $p$ (line 2).
Then,
at line 3,
we do the same in order to compute
the set of blocks $t$ that produced $p$.
In turn,
for every block pair $(b_1, b_2)$
of the Cartesian product $t \times c$,
we check whether there is a happens-before relation
between $b_1$ and $b_2$.
If $b_2$ is not dependent on $b_1$
($b_1 \not\prec_{g} b_2$)
and vice versa,
we report a missing ordering relationship
(line 6).

As a next step,
the algorithm checks for missing notifiers.
If the block $b_2$,
which consumed $p$,
corresponds to a {\tt service} (line 8),
the algorithm examines
whether the relation $b_1 \rightarrow_{g} b_2$ holds
(line 9).
If the block $b_1$,
which produced $p$,
does not send notifications to the {\tt service} $b_2$
the algorithm reports a missing notifier (line 10).

\section{Implementation}
\label{sec:implementation}
Here are our method's
implementation details and
their current limitations.

\vspace{-1mm}

\subsection{Details}

We have developed a prototype that implements
our approach in the OCaml programming
language\footnote{We plan to release it as an open-source software.}.
Our tool consists of three different components:
(1) an executor
that is responsible for
tracing Puppet programs by
taking a Puppet manifest as input
and executes it using {\tt strace} to
collect traces;
(2) an analyzer
that receives a sequence of system calls,
models them in FStrace,
and implements the interpretation rules
presented in Figure~\ref{fig:semantics};
(3) a fault detector for Puppet,
which takes the analyzer's output
and follows the steps of Algorithm~\ref{alg:mor}.
Note that,
we build the dependency graph of a Puppet program
through a simple analysis of the catalogs produced by Puppet
after the compilation of the manifests.
(Catalogs are {\sc json} documents
that list all Puppet abstractions
that are going to be applied along
with their dependencies.)

We have implemented our method with efficiency in mind.
Our tool is able to handle
{\sc gb}-sized traces with reasonable time
and space requirements
(see Section~\ref{sec:eval-performance}).
This was made possible through
a number of optimizations, such as
the use of streams to process and analyze traces,
a reversed {\tt inode} table
to lookup paths based on their {\tt inode}s,
and function memoization.

\vspace{-2mm}

\subsection{Current Limitations}
Currently,
our tool can only support Linux distributions
because {\tt strace} is a utility for
Linux-based operating systems.
However,
we can easily extend it
to support other {\sc posix}-compliant environments
such as FreeBSD or Solaris.
Also,
as we will observe in Section~\ref{sec:evaluation},
our tool might produce false positives
when two Puppet abstractions operate on the same file,
but they are commutative to each other,
i.e., the application order does not matter.
Even though commutative pairs of abstractions
are not so common
(see Section~\ref{sec:evaluation}),
we plan to address this issue in future work
by examining Puppet catalogs
to identify such pairs.

\vspace{-2mm}

\section{Evaluation}
\label{sec:evaluation}

We have evaluated our framework
by examining a large number of Puppet modules
in order to answer
the following research questions.

\begin{enumerate}[label={\bf RQ\arabic*}, leftmargin=2.1\parindent]

\item Is the proposed approach effective
for finding faults in Puppet manifests?
(Section \ref{sec:eval-bugs})

\item How can we categorize the detected faults?
(Section \ref{sec:fault-patterns})

\item What is the performance of our analysis?
(Section \ref{sec:eval-performance})

\end{enumerate}

\vspace{-2mm}

\subsection{Experimental Setup}

We collected a large number
of Puppet modules taken from
Forge {\sc api}\footnote{\url{https://forgeapi.puppetlabs.com/}}
and Github.
We were particularly interested in modules
that support Debian Stretch,
because Debian is one of the most popular
Linux distributions~\cite{linuxdistros}.
We used Docker
to spawn a clean Debian environment efficiently.
Then,
we ran our framework on every module separately.
We monitored the Puppet process
and collected the system call trace of every program
through {\tt strace}.
Finally,
we ran each step
(trace analysis and fault detection)
and logged the reports generated by our framework.
Through this process,
we successfully ran
and analyzed 351 Puppet modules in total.

To compute the performance of our approach
we ran the trace analysis
and fault detection steps ten times
to get reliable measurements.
By examining the standard deviation,
we observed that the running times
did not vary significantly among different executions.
All the experiments were run
on a Virtual Machine with
an 2.1{\sc gh}z 8-core processor
and 8{\sc gb} of {\sc ram}.

\subsection{Fault Detection Results}
\label{sec:eval-bugs}

\begin{table}[t]
\centering
\caption{Faults found in Puppet modules}
\vspace{-2.5mm}
\label{tb:faults}
\resizebox{\linewidth}{!}{%
\begin{tabular}{rlrrrr}
\hline
& & & & & {\bf Fix} \\
{\bf \#} & {\bf Module}                                    & {\bf \# Faults} 
& {\bf MOR}      & {\bf MN}       & {\bf Accepted} \bigstrut \\
\hline
1   &    istlab-stereo                               & 9        & 9        & 0 & \CheckmarkBold\\
2   &    geoffwilliams-auditd                        & 4        & 4        & 0 & -\\
3   &    wiltonms-webserver                          & 4        & 4        & 0 & -\\
4   &    nogueirawash-mysqlserver                    & 3        & 3        & 0 & -\\
5   &    puppet-proxysql                             & 3        & 3        & 0 & \CheckmarkBold\\
6   &    saz-ntp                                     & 3        & 1        & 2 & -\\
7   &    deric-mesos                                 & 3        & 0        & 3 & \CheckmarkBold\\
8   &    hardening-os\_hardening                     & 2        & 2        & 0 & \CheckmarkBold\\
9   &    saz-locales                                 & 2        & 2        & 0 & -\\
10  &    vpgrp-influxdbrelay                         & 2        & 2        & 0 & \CheckmarkBold\\
11  &    jgazeley-freeradius                         & 2        & 1        & 1 & \CheckmarkBold\\
12  &    ploperations-puppet                         & 2        & 1        & 1 & -\\
13  &    walkamongus-codedeploy                      & 1        & 1        & 0 & \CheckmarkBold\\
14  &    spynappels-support\_sysstat                 & 1        & 1        & 0 & -\\
15  &    cirrax-dovecot                              & 1        & 1        & 0 & \CheckmarkBold\\
16  &    sgnl05-sssd                                 & 1        & 1        & 0 & -\\
17  &    roshan-mysqlzrm                             & 1        & 1        & 0 & -\\
18  &    puppetfinland-nano                          & 1        & 1        & 0 & \CheckmarkBold\\
19  &    olivierHa-influxdb                          & 1        & 1        & 0 & -\\
20  &    noerdisch-codeception                       & 1        & 1        & 0 & -\\
21  &    nextrevision-flowtools                      & 1        & 1        & 0 & \CheckmarkBold\\
22  &    baldurmen-plymouth                          & 1        & 1        & 0 & \CheckmarkBold\\
23  &    alertlogic-al\_agents                       & 1        & 1        & 0 & -\\
24  &    puppet-telegraf                             & 1        & 0        & 1 & \CheckmarkBold\\
25  &    puppetlabs-apache                           & 1        & 0        & 1 & \CheckmarkBold\\
26  &    example42-apache                            & 1        & 0        & 1 & \CheckmarkBold\\
27  &    deric-zookeeper                             & 1        & 0        & 1 & \CheckmarkBold\\
28  &    camptocamp-tomcat                           & 1        & 0        & 1 & -\\
29  &    alexharvey-disable\_transparent\_hugepage   & 1        & 0        & 1 & -\\
30  &    camptocamp-ssh                              & 1        & 0        & 1 & -\CheckmarkBold\\
\hline
& {\bf Total}                                    & {\bf 57} & {\bf 43} & {\bf 14} & - \bigstrut \\
\hline
\end{tabular}}
\vspace{-4mm}
\end{table}

Our framework detected
57 previously unknown issues in 30 Puppet modules.
Table~\ref{tb:faults} presents
the analysis results for each module.
Notably,
this is the first study
that led to the disclosure of
such a large number of faults in Puppet repositories.
Our framework marks 43 out of 57 faults
as missing ordering relationships (column {\sc mor}).
We observe that ordering violations
are the most prevalent issue
in the inspected Puppet manifests.
The rest of the faults
are related to missing notifiers (column {\sc mn}).

Based on the reports of our tool,
we manually verified
that each reported fault can lead to a problematic situation
by reproducing each case.
We provided fixes for 21 projects,
and 16 of them were accepted by their development teams and
integrated into their code.
This indicates that our tool
produces reports that are meaningful to developers.
At the time of the submission,
none of our patches have been rejected.

\subsection{Fault Patterns}
\label{sec:fault-patterns}

Below,
we categorize and discuss
some of the faults identified by our framework.
Most represent previously unknown to us fault patterns
which we learned through our tool. 

\subsubsection{Missing Ordering Relationships}
We have observed two  types of
missing ordering relationships issues.

{\bf Generate-Use Violation.}
The use of a resource
must always succeed its creation.
Many modules fail to preserve
that ordering relationship.
We observed this violation in 16 Puppet modules
such as {\tt alertlogic-al_agents},
{\tt hardening-os\_hardening}, etc.
Figure~\ref{fig:fault1} shows
a fragment from
{\tt alertlogic-al_agents}~\cite{alertlogic}.
The code first fetches a {\tt .deb} package
(a Debian archive)
using the {\tt wget} command
(lines 1--5).
The package is stored at the path
specified by the {\tt \$package_path} variable
whose value is {\tt /tmp/al-agent}.
Then,
the code installs the Debian archive
on the system (lines 6--10)
through {\tt dpkg}.
\footnote{{\tt dpkg} is a package management system for
Debian-based operating systems}
It is easy to see
that the {\tt package} depends on {\tt exec}
because it requires {\tt \$package_path}
(the {\tt .deb} file) to exist in the system (line 9)
so that it can install the package successfully.

The \emph{Generate-Use} category
produces observable errors
(errors that manifest during
the catalog's application),
when Puppet applies abstractions in the erroneous order.
For example,
when it processes {\tt package} before {\tt exec}
the application of the catalog fails
with the following error:
``{\tt dpkg: error: cannot access archive "/tmp/al-agent": No such file or directory}''

\begin{figure}[t]
\begin{lstlisting}[language=puppet]
$package_path = "/tmp/al-agent"
exec {"download":
  command => "/usr/bin/wget -O ${package_path} ${pkg_url}",
  creates => $package_path
}
package {"al-agent":
  ensure   => "installed",
  provider => "dpkg",
  source   => $package_path,
}
\end{lstlisting}
\vspace{-3mm}
\caption{{A Missing Ordering Relationship between
{\tt package} and {\tt exec}}.}
\label{fig:fault1}
\vspace{-5.5mm}
\end{figure}

{\bf Configure-Use Violation.}
The configuration of a file
must precede its use.
For example,
when a service starts,
all the files consumed by that service
have to be properly configured.
This category differs from
the previous one
because when a Puppet abstraction attempts to use the file,
the latter exists in the system.
However,
this is not in the expected state
(e.g., the file does not have the right contents,
permissions, etc).
This error pattern
appears in four modules:
{\tt saz-ntp},
{\tt vpgrp-influxdbrelay},
{\tt olivierHa-influxdb},
and
{\tt ploperations-puppet}.

Figure~\ref{fig:motivation-ex1}
illustrates a program with
an issue related to this category.
When the {\sc ntp} service starts,
the configuration files are guaranteed to be there
because the abstraction {\tt package} creates them
during installation.
However,,
it is possible that the {\sc ntp} service does not read
the desired contents of the {\tt /etc/default/ntp} file
specified by {\tt content => "conf content..."}
(line 11),
because there is a missing ordering relationship
between {\tt file} and {\tt service}.
Note that this category---unlike
the previous one---might
lead to unexpected behaviors silently,
i.e., the application of the catalog
does not produce any error messages.

\vspace{-0.5mm}
\subsubsection{Missing Notifiers}

We have identified four different categories
of issues related to notifiers.

{\bf Configuration Files.}
A configuration file must always send
notifications to a service
so that any change to that file triggers
the restart of the corresponding service.
Although this is a standard pattern,
we observed that in four modules
(shown in rows 6, 7, 11, 12, 30 of Table~\ref{tb:faults})
this is not the case.
As an example recall the program
discussed in Section~\ref{sec:overview}.

{\bf Log Files.}
Typically,
services log various events
in dedicated files.
For instance,
the log file of an Apache server records---among
other things---every incoming {\sc http} request.
Log files are very beneficial
for debugging and monitoring purposes.
When a service starts,
it opens a corresponding log file,
which remains open,
while the service is up,
to write any events
that take place.

We discovered issues related to logging
in two popular Puppet modules
({\tt puppetlabs-apache}~\cite{puppetlabs-apache},
and {\tt deric-zookeeper}~\cite{zookeeper}).
These modules declare the log files for
the {\tt apache} and {\tt zookeeper}
services in their manifests.
However,
the log files do not have a notifier
for their associated services.
This may lead to a problematic situation.
Consider the case where the log file of a service
is removed from the host.
When we remove an open file,
the underlying system call
({\tt unlink})
does not update the file descriptors
associated with the removed file,
even though Puppet will create a new one.
This means that
although the file disappears from the file system,
the service still handles a file descriptor
that points to the {\tt inode} of the original file.
The issue is that
after removal,
the {\tt inode} becomes an \emph{orphan}
(i.e., it is not linked with any file),
which means that it is no longer accessible
through a file path.
Therefore,
in the case of a missing notifier,
the log history of the upcoming events is lost
because the service writes to an orphan {\tt inode}.
To fix that issue,
the log file should notify the service
so that the service opens the
newly-created log file.

\begin{figure}
\begin{lstlisting}
# config.pp manifest
define tomcat::instance::config (..., $basedir, $javahome) {
  file {"/etc/init.d/tomcat-${name}":
    ensure  => $present,
    content => template("tomcat/tomcat.init.erb"),
    require => Concat["${catalina_base}/bin/setenv.sh"],
  }
}
# tomcat/tomcat.init.erb template
export JAVA_HOME=<%= @javahome %>
export CATALINA_BASE=<%= @basedir %>
export CATALINA_PID=<%= @basedir %>/temp/tomcat.pid
\end{lstlisting}
\vspace{-4.5mm}
\caption{Manifest and its template taken from {\tt camptocamp-tomcat}.
We omit irrelevant code for brevity.}
\label{fig:tomcat-ex}
\vspace{-5mm}
\end{figure}

{\bf Init Scripts.}
Init scripts
specify how a service starts,
stops or restarts.
In practice,
they are wrapper shell scripts
which setup the required environment
and invoke the actual executables of the services
with the appropriate parameters.

Puppet manifests,
that manage init scripts
should notify the corresponding service
whenever there is a change to those scripts.
The {\tt camptocamp-tomcat}~\cite{tomcat}
and
{\tt alexharvey-disable_transparent_hugepage}~\cite{alex}
modules fail to follow that pattern.
Consider the code listed in Figure~\ref{fig:tomcat-ex},
coming from the {\tt camptocamp-tomcat} module.
The Figure shows fragments,
coming from two different files.
First,
the {\tt config.pp} Puppet manifest,
defines a custom abstraction named {\tt tomcat::instance::config},
which takes the variables {\tt \$basedir}
and {\tt \$javahome} as parameters (line 1).
This abstraction
configures the init script of the tomcat service (lines 2--8)
whose contents are determined
by the {\tt tomcat/tomcat.init.erb} template (lines 9--12).
By examining this template we see that,
before the init script starts tomcat,
it sets some environment variables
based on the values of the Puppet parameters
{\tt \$basedir} and {\tt javahome} (lines 10--12).
When there is an update to the init script
(e.g., {\tt \$javahome} variable has a different value),
Tomcat should restart
in order to operate on the new environment,
e.g., to use a different version of Java.

{\bf Packages.}
When Puppet applies a {\tt package} abstraction,
the service that depends on that package should restart.
In this way,
we ensure that
a service gets all the necessary updates,
including, security patches, new features, etc.
Our tool identified this kind of issue
in {\tt example42-apache}~\cite{example42},
{\tt saz-ntp}~\cite{ntp},
and {\tt puppet-telegraf}~\cite{telegraf}.
Specifically,
the {\tt package} abstractions
that were responsible for installing
the Apache, {\sc ntp}, and telegraf
did not notify the running instances
whenever there was a new version of those packages.

\subsection{Performance Evaluation}
\label{sec:eval-performance}

\begin{figure}[t]
\includegraphics[scale=0.33]{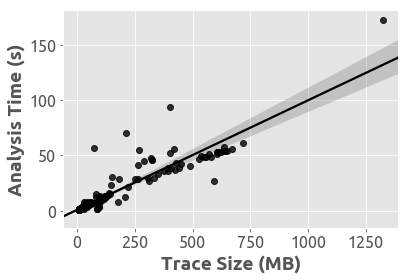}
\vspace{-2.5mm}
\caption{The trace analysis
and fault detection time as a function of the trace size.
Each spot shows the average time spent on
both the trace analysis and fault detection phases
for a given trace obtained by the execution of a module.}
\label{fig:performance}
\vspace{-6mm}
\end{figure}

Figure~\ref{fig:performance} shows
the running times (in seconds) of the
trace analysis and fault detection phases
relatively to the size of the provided traces
(in {\sc mb}).
We observe
that the correlation between the trace size
and analysis time is almost linear.
Notice that our framework
is able to handle a large volume of traces
(more than 1.2{\sc gb})
in a reasonable amount of time ($<$ 3 minutes).
The average trace size
and analysis time of the inspected modules
is 84{\sc mb} and 9 seconds respectively.

There are 4 cases out of 351
({\tt ceritsc-dkms},
{\tt datadog-datadog_agent},
{\tt nexcess-ksplice},
{\tt puppet-rabbitmq})
where the execution times were relatively high
compared to the rest of the modules.
Nevertheless,
they all remain in acceptable limits.
By examining the characteristics of the traces
obtained by the execution of these modules,
we observed that they contain more
{\tt unlink} system calls than the rest of the modules.
Notably,
such calls involve more expensive operations
on the analysis state
(they are modeled as \prim{hpath} $d, p$ \prim{expunged},
recall Figure~\ref{fig:semantics}).

Overall, the overhead of the analysis is relatively small.
We argue that our approach is practical
and can be used as part of the testing process
for Puppet manifests.

\subsection{False Positives}

We have manually inspected the reported issues
and identified a potential source of false positives.
Consider two abstractions that are
commutative to each other.
For example,
in the {\tt claranet-varnish} module~\cite{varnish},
the developers use two different abstractions
to partially configure a certain file.
On the one hand they use {\tt file} to set the permissions
and ownership of the file,
and on the other they use {\tt exec} to
initialize its contents.
In this case the execution order
in which Puppet processes abstractions
does not matter.
Specifically,
Puppet can first use {\tt exec}
to create the file with the desired contents,
and then apply {\tt file} to set
the appropriate file's attributes
or vice versa.
Our approach reported false positives
\emph{only} in 7 
out of 351 cases.
Therefore,
we argue that this pattern
(i.e, configuring a file through the
combination of abstractions)
is not particularly common.

We noticed one more false positive
which was related to missing notifiers.
The developers of
{\tt bodgit-dbus}~\cite{bodgit}
use a custom command (expressed via {\tt exec})
to reload the configuration of the service.
Consequently,
the configuration files notify the {\tt exec} abstraction
instead of {\tt service}.
We did not observe this case elsewhere,
because Puppet programmers typically exploit
the {\tt restart} parameter
of the {\tt service} type
to define a custom restart command
in the following manner:
``{\tt service \{ restart => "/custom/cmd", \ldots \}}''.


\vspace{-1.5mm}
\section{Related Work}
\label{sec:related}

Our work is related to three research areas,
namely quality in IaC,
trace analysis,
and modeling of file systems operations.

{\bf Quality in IaC.}
With the proliferation of
the IaC process,
there have been numerous attempts to
identify defects
and quality concerns in configuration code.

A number of studies focus on
maintainability issues.
Sharma et al.~\cite{sharma2016}
design and implement a code-smell detection
scheme for Puppet,
which searches for issues
related to naming conventions,
code design, indentation, etc.
Their findings suggest
that such anti-patterns---as in the traditional programs---exist
in many IaC repositories.
Van der Bent et al.~\cite{gousios2018}
introduce a quality model for Puppet programs
which is empirically evaluated
by interviewing practitioners from industry.
Schwarz et al.~\cite{chefsmells}
do similar work focusing on Chef recipes.
Endeavors have recently moved to
the identification of security issues.
Rahman et al.~\cite{security}
define and classify security smells into seven categories,
such as hard-coded passwords and the
use of weak cryptographic algorithms,
and then build a tool for
statically detecting these smells
in Puppet repositories.

Other studies attempt to
extract error patterns
and source code properties from
the analysis of defective IaC programs.
Rahman et al.~\cite{rahman2018char}
employ machine learning
and text processing techniques
to identify properties
that faulty Puppet programs hold.
Then,
they build a prediction model
for asserting whether IaC scripts manifest faults or not.
Chen et al.~\cite{chen2018} identify
error patterns in Puppet manifests
by following a different approach.
First,
they inspect the code changes from
repositories' commits.
Second,
they construct an unsupervised learning model
to detect error patterns based on
the clustering of the proposed fixes.
Their approach is based on the assumption
that similar faults are fixed with similar patches~\cite{hanam2016}.

There are few automated techniques
proposed for improving the reliability of
configuration management programs.
Rehearsal~\cite{rehearsal}
statically verifies that a given Puppet configuration
is deterministic and idempotent.
Rehearsal models a given Puppet manifest
in a small language called {\sc fs}
and then it constructs logical formulas
based on the semantics of each language's primitive.
Finally,
an {\sc smt} solver decides
whether the initial program is non-deterministic or not. 
Compared to our approach,
Rehearsal is less effective and practical.
Specifically,
Rehearsal employs a form of static analysis
that can only handle a subset of Puppet programs.
For example,
the analysis does not support {\tt exec} abstractions
because it cannot reason about
the file system resources
that {\tt exec} processes.
Unlike Rehearsal,
our approach operates on the actual system calls
rather than Puppet manifests;
thus,
it can effectively determine
which files are affected
by a Puppet run and how.

Other advances~\cite{testing,citac} adopt
a model-based testing approach
for checking whether configuration scripts meet
certain properties.
Hummer et al.~\cite{testing}
focus on testing the idempotence of Chef scripts.
Their proposed framework generates multiple test cases
that explore different task schedules.
By tracking the changes in the system
(they compare the system state before and after execution),
they determine if idempotence holds for a given program.
Hanappi et al.~\cite{citac} extend the work of Hummer et al.
and introduce Citac;
a framework that can be applied to Puppet manifests
to examine the convergence of programs.
Convergence states
that the system reaches a desired state
even at the presence of failed Puppet abstractions.
They formally express
the properties of idempotence and convergence,
and through test case generation,
they verify if the provided manifests
violate those properties.
Contrary to Citac,
we adopt a more lightweight and practical approach
applying manifests only once.
Finally,
neither Rehearsal nor Citac detect
issues involving missing notifiers.

{\bf Trace Analysis.}
Analysis of system call traces
has been widely used in the past,
especially for
intrusion and malware detection~\cite{intrusion,intrusion2,malware,malware2}. 
Mutlu et al.~\cite{mutlu}
collect execution traces from JavaScript applications.
Their traces do not track system calls,
but they capture memory and storage (e.g., cookies) accesses
in the context of the browser.
They split traces into blocks,
where each block describes
the execution of an asynchronous callback
(e.g., {\sc ajax} handler).
As the execution of each handler is partially ordered,
they apply a simple data-flow analysis over traces
to join the states coming from different handlers.
In this manner,
they effectively detect data races
by identifying handler pairs
where the merges of their corresponding states
result in different values of the same variable.
In our work,
we also separate the trace sequences into blocks.
However,
we are interested in file system operations
instead of reads
and writes to memory locations.
Also,
we apply a different methodology
for discovering execution blocks
that might lead to harmful scenarios.

{\bf Modeling File System Operations.}
Several researchers have designed
specifications for the {\sc posix} file system~\cite{hesselink,freitas,gardner}.
The specifications mainly focus on program reasoning
and verification.
Furthermore,
Shambaugh et al.~\cite{rehearsal}
have introduced {\sc fs};\@
a small language used
to model the effects of Puppet abstractions on
the file system.
In this work,
we model system calls rather than Puppet abstractions.

\vspace{-2.5mm}
\section{Conclusion}
\label{sec:conclusions}

We have introduced a novel and practical approach
for identifying missing dependencies
and notifiers in Puppet programs.
Our method collects the system calls invoked by
a Puppet program and models them in FStrace.
Through FStrace,
we capture how higher-level programming constructs,
such as Puppet abstractions,
interact with the operating system and derive
their relationships.
Then,
our method checks the inferred relationships
against the program's dependency graph
and reports potential mismatches.

The effectiveness of our approach
is exemplified by the uncovering
of 57 previously unknown issues in 30 projects.
Notably,
we provided fixes for 21 modules
and 16 of them were accepted by the developers.
We have further showed
that our tool can handle realistic traces
in a reasonable time.
Our results indicate that our tool can be used
as part of the testing process for Puppet programs.

FStrace is a generic model that
can be applied to other domains
with partially ordered constructs
such as the asynchronous callbacks of JavaScript.
Recent work~\cite{nodejs,nodefuzz} has showed
that many concurrency faults in Node.js applications
are caused by data races that appear in files
instead of memory locations.
As future work,
we are planning to leverage our method
to detect such concurrency faults in
JavaScript server-side applications.

\bibliographystyle{ACM-Reference-Format}
\bibliography{main}

\end{document}